\documentclass{aa} 
\usepackage{epsfig,natbib} 
\newcommand{\sqd}{\mbox{\rm\ sq. deg.}} 
\def\msun{{\rm\,M_\odot}} 
\begin{document} 
%
\title{Dark baryons not in ancient halo white dwarfs
\thanks{Based on observations made at
Canada-France-Hawaii Telescope (CFHT)}}

   \author{Michel Cr\'ez\'e
        \inst{1,8}
        \and
        Vijay Mohan \inst{3}
        \and
        Annie C. Robin \inst{2}
        \and
        C\'eline Reyl\'e \inst{2}
        \and
        Henry J. Mc Cracken \inst{6}
        \and
        Jean-Charles Cuillandre \inst{4}
        \and
        Olivier Le F\`evre \inst{5}
        \and
        Yannick Mellier \inst{6,7}
         }

 \institute{ 
    Universit\'e de Bretagne Sud, BP 573 
    F-56017 Vannes Cedex, France 
    \and 
    CNRS UMR6091, Observatoire de Besan{\c c}on, BP1615,  
    F-25010 Besan\c{c}on Cedex, France
    \and 
    U.P.  State Observatory, Manora Peak, 
    Nainital, 263129 India
    \and 
    C.F.H.T. Corp. P.O. box 1597, Kamuela Hawaii 96783, USA
    \and 
    Laboratoire d'astrophysique de Marseille,Observatoire
    de Marseille Provence, Universit\'e de Provence et CNRS,
    BP8 traverse du Siphon, 13376 Marseille Cedex France
    \and  
    Institut d'Astrophysique de Paris, 98 bis Boulevard Arago, 75014 
    Paris , France
    \and  
    LERMA, Observatoire de Paris, 61 avenue de l'observatoire, 75014 
    Paris, France 
    \and
    PCC, Coll\`ege de France, 11 place Marcellin Berthelot, 75231
    Paris Cedex 05
          }
\offprints{Michel Cr\'ez\'e, \email{Michel.Creze@univ-ubs.fr}}

\date{Received ; accepted }

\abstract{
Having ruled out the possibility that stellar objects are the main contributor 
of the dark matter embedding galaxies, microlensing experiments
cannot exclude the 
hypothesis that a significant fraction of the Milky Way dark halo might be 
made of MACHOs with masses in the range $0.5-0.8 \msun$. Ancient white 
dwarfs are generally considered the most  plausible candidates for such 
MACHOs. We report the results of a search for such white dwarfs in a proper 
motion survey covering a $0.16\sqd$ field at three epochs at high galactic
 latitude, and  $0.938\sqd$ at two epochs  at intermediate galactic 
 latitude (VIRMOS survey), using the CFH telescope. Both surveys are complete 
 to $I = 23$, with detection efficiency fading to 0 at $I = 24.2$. Proper 
 motion data are suitable to separate unambiguously halo white dwarfs 
 identified as belonging to a non rotating system. No candidates were found 
 within the colour-magnitude-proper motion volume where such objects can be 
 safely discriminated from any standard population as well as from possible 
 artefacts. In the same volume, we estimate the maximum white dwarf halo
 fraction compatible with this observation at different significance
 levels if the halo is at least 14 gigayears old  and under different
 ad hoc initial mass 
 functions. Our data alone rule out a halo fraction greater than $14 \% $  
 at a $95\% $ confidence level. Combined with two previous investigations
 exploring comparable volumes ,this pushes the limit below $4 \% $ ($95\% $ 
 confidence level) or below $1 \% $ ( $64\%$ confidence), and implies
 that if baryonic dark matter is present in galaxy halos, it is not,
 or is only
 marginally in the form of faint hydrogen white dwarfs.
}

\maketitle

\section{Introduction} 
\maketitle

%
 
A non rotating or weakly rotating, temperature supported halo is what
seems to best satisfy most existing requirements for the bulk of dark
matter implied by the large-scale rotation curve of the Milky Way
galaxy. If a more than marginal part of this dark matter is made of
baryons, then it should be expected to appear in the form of
moderately massive compact objects.  
   
Concerning the identity of such objects, there were two kinds of
candidates initially envisaged : Jupiter-like objects or brown dwarfs
on  the low mass side, very old fading degenerate stars (White Dwarfs)
 at larger masses $ (0.5-0.8\ \msun)$. Brown Dwarfs and planets
were eliminated by early microlensing experiments
\citep{alcock,aubourg}. Therefore cool
white dwarfs appeared as the only possibility left  
\citep{charlot}. \cite{Chabrier99} and \cite{Chabrier2000b}
show that ad hoc star formation scenarios are able to form an
adequate number of such objects out of the protogalactic matter
without producing unrealistic heavy element enrichment.   
 
Second generation microlensing surveys at high galactic latitude rule 
out heavy halos fully made of very old white dwarfs
\citep{aubourg}. It is still not 
clear whether they imply some fraction of the halo in the form 
of massive compact halo objects (MACHOS). The EROS collaboration 
\citep{afonso2003} gives only an upper limit of about 25\% for
masses ranging from  $0.5$ to $0.8~\msun$. The MACHO 
collaboration \citep{alcock} gives a range [20\%,50\%] which seems to exclude  
fully non-baryonic halos, yet they admit that this result is somewhat 
model-dependent.

Direct detections of very ancient white dwarfs in colour-proper motion  
surveys have now been obtained by different groups \citep{Oppenheimer, 
ibata2000,Ruis2001,Mendez2002}, several of them with
spectroscopic confirmation. However, in all positive cases, the sampling
conditions were such that it is not possible to identify candidates
unambiguously connected to the non-rotating component. Detected
ancient white dwarfs are most likely members of the thick disc of 
the Milky Way \citep{Reyle2001b,Reid,Torres2002}. 
 
In contrast, deeper surveys explore regions   
of the colour-, magnitude proper motion space where halo white dwarfs 
are the only expected contributors.
 In such surveys \cite{ibata}, \cite{nelson}, \cite{goldman} do not 
report detections in these
regions. The detections reported by \cite{nelson} are not in the 
proper motion range where there is no ambiguity.
Also, attempts to search for similar objects in Luyten's LHS and
NLTT large proper motion surveys \citep{flynn2001} found none.

We report here on a ground-based effort to detect ancient 
halo white dwarfs from a deep photometry-proper motion 
survey (section~2). The volume explored is 3 times that of the 
Nelson's field, almost the same order of magnitude as Goldman's. A  
colour-magnitude-proper motion criterion is 
established, isolating a volume where, based on a realistic range of 
galaxy models, the expected contributions of disc and thick disc 
components are negligible, while the detection efficiency is quite 
high. In this volume the expected contribution of halo 
models made of ancient white dwarfs is evaluated (section~3). An
estimation of the halo fraction in the form of cool hydrogen white
dwarfs compatible with getting no detection in the observed sample is
performed. Similar  model predictions are produced for the observing
conditions of previous investigators in order to produce a combined
constraint on the halo fraction (section~4 and conclusion).

 
%
%
 
\section{Survey description } 
 
 
Observational data used in this investigation come from two  
different observing programs using wide field CCD cameras at the 
3.60 meter CFH Telescope at Mauna Kea (Hawaii).
One program near the galactic pole (herein with SA57 field) 
in special selected area SA57 is the result of three observation 
campaigns by some of the authors of this paper. 
The other one is part of the so-called VIRMOS survey
\citep{lefevre2004,McCracken2003}. Both are almost equally deep.  
 
%
%
 
\subsection{Field characteristics}

 
\subsubsection{VIRMOS field} 
         
The VIRMOS survey uses the CFH12K camera. Out of the full VIRMOS 
survey fields, one subfield lying in the galactic
anticenter direction  
($\alpha_{2000} = 2^{h} 26^{m}, \delta_{2000} = -4^{o} 30^{'}$;  
$l^{II} = 172^{o}, b^{II} = -58^{o}$) 
was observed on two epochs in 1999 and 2000 in  V and I pass bands. The 
useful sky area covered is 0.938 sq. degree. 
 
 
\subsubsection{SA57 field} 
 
 One north galactic pole field was observed in Selected Area SA57($\alpha$=13h
9m 5s, $\delta$=+29$^{o}$ 9' 30''). Observations were obtained
 during three observing runs (March
21-23 1996, April 23-25 1998, May 27-29 2000 ). The UH8K camera, a
mosaic of 8 CCDs each of 2048x4096 pixels was used during
the first two runs. The CFH12K was used on the third run. In  both cameras,
the projected pixel size is 0.205 arcsec wide on the sky. The useful field is
limited to the overlap between the three observing epochs, that is
0.16 square degrees. Series of  20 minute exposures were obtained in the
V, R and I filters. The total exposure time is 2 hours in each filter at each
epoch.
 
The seeing was better than $0.7$ arcsec in 1998, about $0.9$ in 1996 and
about 0.8  
in 2000. The 1998 run gives the best tool for star-galaxy separation and for  
faint object detection confirmation, but the completeness limit for proper  
motions is imposed by the 1996 run. In these conditions the
completeness limit turns out to be somewhat below $25$ in V for extended
sources, as seen from the magnitude histogram. Completeness is
expected slightly deeper for star-like   
sources. Full discussion of the detection efficiency issue for star like  
objects is given in sections 2.5 and 2.6. 
 
%
%
%
\begin{figure} 
\epsfig{file=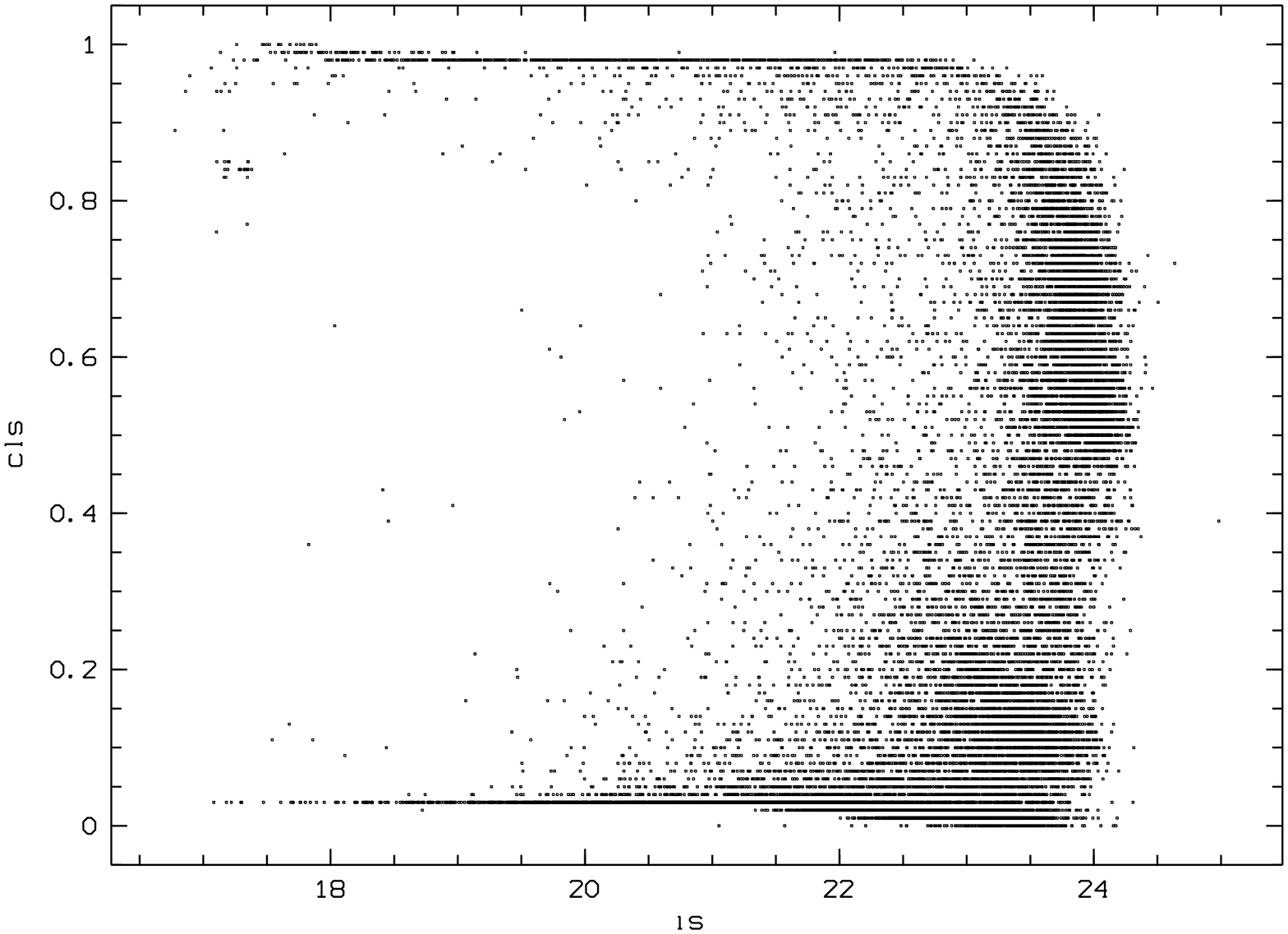,width=6cm,angle=270} 
\epsfig{file=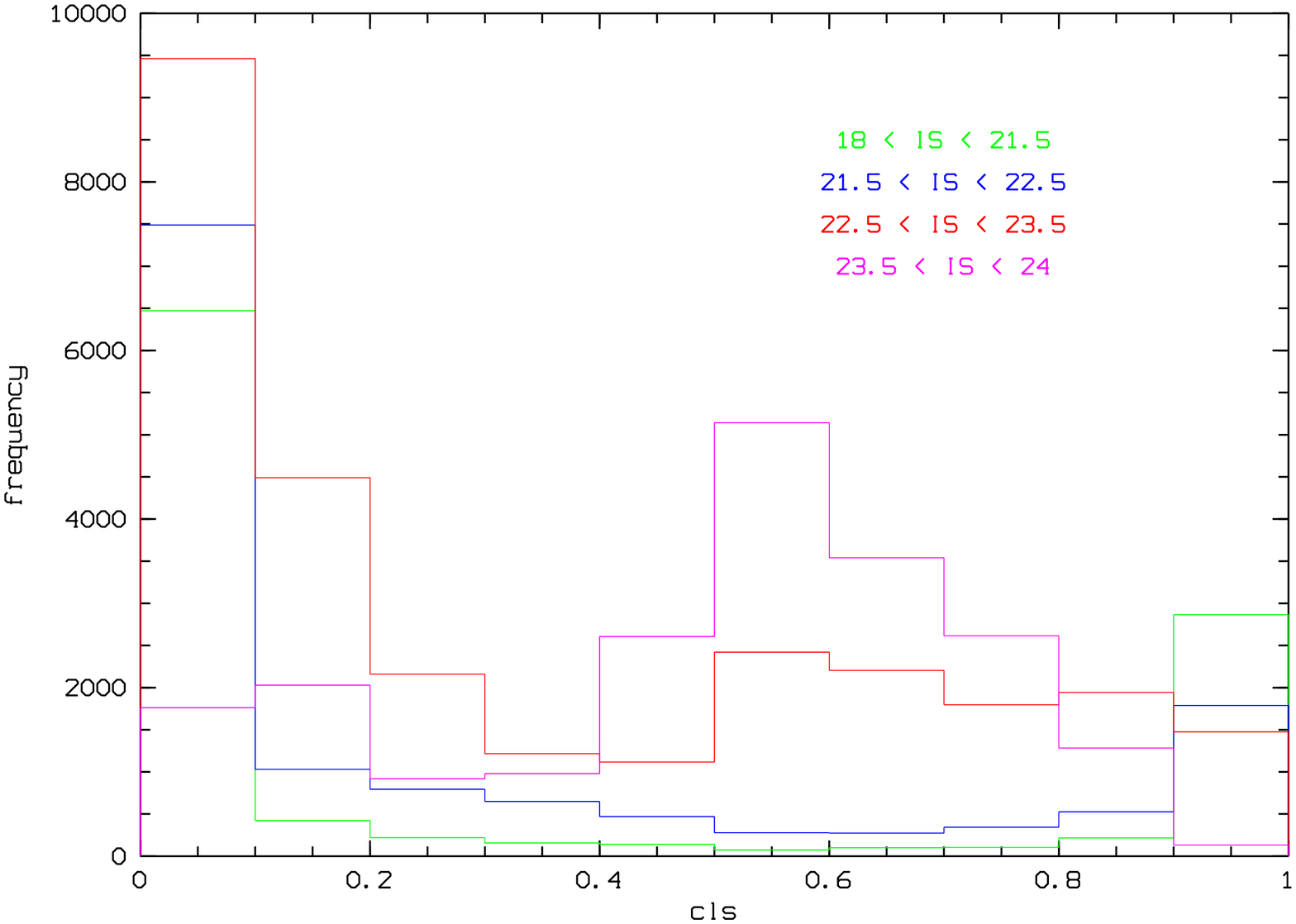,width=6cm,angle=270} 
\caption { Star galaxy separation : the CLS classifier runs from 0  
(safely non-stellar) to 1 (safely stellar). Top : CLS is versus I magnitude
(in the Cousin's system). Bottom :  CLS frequency distributions in 4
magnitude ranges. Beyond magnitude 23.5 the galaxy peak at $CLS < 0.1$
vanishes, indicating that the classifier no longer recognizes them.} 
\label{cls-mag}  
\end{figure} 
 
\subsection{Source extraction and classification} 
  
Source extraction was performed on the sum of all images at each epoch
using   
the SExtractor package \citep{bertin}. Associated with the  
source extraction, SExtractor provides a classification parameter named  
 ``class star" (CLS) ranging from 0 (safely non-stellar) to 1 (safely
star-like).  
This class is computed by a neural network which uses the distribution
of flux   
between object pixels as input data and a library of simulated star
and galaxy  
images as training set. Rather than replaying the training with a specific  
library, we preferred to cross check and calibrate the diagnostic on real  
images.

 The resulting distribution of $CLS$ as a function of magnitude is
 plotted in Fig.~\ref{cls-mag}, the VIRMOS field has been chosen for
 illustration. It is obvious from the histograms in the
 lower panel that at magnitudes brighter than $I=22.5$, the
 contamination of the high $CLS$ range by non-stellar objects is very
 marginal because there are not many galaxies.

The effect of the CLS selection upon detection efficiency is
  investigated in section~2.5. For deriving the accuracy of proper
  motions, (section~2.3) non-stellar sources would heavily affect 
the statistics, hence we adopt a high ($CSL > 0.8$) selection 
threshold. On the contrary, when it comes to detecting as large a
number of white dwarf candidates as possible, the detection efficiency
 must be maximum.
Then the $CLS$ selection  
 threshold for searching halo white dwarfs (section 3), which in the
 present case amounts to delimitating the volume within which there
 are no white dwarfs detected, has been set to 0.4. According to
 Fig.~\ref{cls-mag} this threshold
 is sufficient to eliminate the overwhelming majority of obviously non
 stellar objects. Then we rely upon the proper motion
 criterion to eliminate misclassified galaxies.
 

\subsection{Astrometry} 
 
While the SEXtractor package turns out to be an efficient source
detection and classification tool, extensive tests and comparisons
showed that the performances of the photometry and centering algorithm
are not optimal for star-like sources. So,
the DAOPHOT/ALLSTAR package \citep{Stetson1987} was used to estimate 
 magnitudes and relative ($x,y$) positions on each CCD, in each band,
 at each  epoch. Photometry and astrometry are obtained by fitting a variable
empirical PSF separately in each CCD. It was also investigated whether the
goodness of fit parameters provided by DAOPHOT would bring 
additional information concerning the star/galaxy separation. These
attempts were not conclusive. The goodness of fit parameters are
partially redundant with the neural network approach, with less
discrimination power.

\begin{table}  
      \caption[]{VIRMOS field, estimated standard errors of dx and dy differences
        (pixels) as a function of I magnitude. Ns is the number of
        sources used in the statistics}   
         \label{dxdystderr} 
\begin{center}
\begin{tabular}{c r r r r r}   
\hline\hline 
\noalign{\smallskip}
 I mag   & 18-20    &    20-21    &    21-22   &  22-22.5   &  22.5-23
 \\ 
\noalign{\smallskip}
\hline 
\noalign{\smallskip}
Ns   &    951    &      791   &        994  &          512  &    486 \\ 
 
$\sigma_{dx}$ &   0.045   &     0.040  &      0.058   &      0.095   &  0.120 \\  
        
$\sigma_{dy}$ &   0.037   &     0.042  &      0.059   &      0.093   &
0.110 \\
\noalign{\smallskip}
\hline 
\end{tabular}
\end{center} 
\end{table} 
   
The astrometric accuracy can be assessed by the statistics of $x$ and $y$  
differences between two epochs after correction of systematic effects related  
to colour and position detected in the VIRMOS field (see below). 
 
To perform the statistics, objects with starlike profile at $CLS \ge
0.8 $  have  been selected to avoid accuracy degradation related to extended   
sources. Then the sample was split into 5 magnitude ranges each containing 
500 sources or more.  
In order to avoid contamination of the statistics by real significant proper  
motions, a robust statistics based on the central part of the histogram is  
derived from probability plots. The resulting standard errors are
given in Table~\ref{dxdystderr}.  
 
For all useful purposes the trend of dx, dy s.e. was approximated 
by the following formula \textbf{(in pixel units)} : 
\begin{equation} 
  \sigma_{dx} = 0.04 +0.05* \exp(1.094*I_{mag}-24.0)
\end{equation}

 
\subsubsection{VIRMOS proper motion data}  
         
In the case of VIRMOS survey, the pipeline processing from which we got
the data includes re-coordination of all the frames at each epoch in a single absolute  
astrometric reference frame. As a consequence, the reference to individual  
CCD pixels was lost. This turned out to generate various disturbances, each 
of little significance for the large scale galaxy survey which was the main  
goal of the VIRMOS collaboration. However for our proper motion purposes each 
effect deserved specific care. 
 
\begin{itemize} 
 
\item Due to the offset strategy between exposures we had to trace approximately 
 regions of full overlap between individual CCDs in the composite
 epoch image. 
 This restriction is responsible for reducing the proper motion survey from  
 the nominal $1.2\sqd$ down to an effective $0.938\sqd$ 
  
 \item Individual relative $(x,y)$ positions, as remeasured by the DAOPHOT  
 PSF centering algorithm on each epoch image, reflected the possible
 distortions of the global reference frame. Distortion has been
 corrected a posteriori. We have obtained an estimation of the necessary correction by 
assuming that the average motion of a subsample dominated by remote 
objects is zero everywhere in the field. The estimation has been
performed twice by computing the regressions
$ \Delta~dx  = dx (x,y)$ 
at order 3 in both $dx$ and $dy$. This was done separately~: 

a) On a subset of galaxy like  objects selected on the basis having low CLS
but not being too extended.  

b) On a selection of high CLS sources with selection $19 < I < 22$.
 
The pattern of systematic effects obtained from either data sets is quite the
same. For the final data we choose the regression based on stellar
image which is less dispersed. Proper motions were then corrected for
these effects. The maximum correction across the field is less than
 $0.2$ pixels in both $dx$ and $dy$. So it is not likely to disturb
large proper motion investigations.

\item Another distortion appears as a function of colour. It turns out  
to affect $dx$ proper motion only. It is linear in V-I, its amplitude  
is about $0.15$ pixels across the whole colour range. It also affects  
positions of galaxies as well as stellar sources. The effect has been corrected
 following exactly the same process as for the $x,y$
correction.  
 
\end{itemize} 
 
Due to its position in the anticenter direction, proper motion in longitude 
coincides almost exactly with the rotation component.

\subsubsection{SA57 proper motion data} 
 
Relative proper motions are obtained by cross-matching DAOPHOT
positions separately within each CCD between two epochs and
independently in the 3 bands.  Systematic displacements and
distortions between two epochs are modelled by a second degree
polynomial in x and y, including cross term,  for the 96-98 period. A
third order polynomial has been necessary to cross-match 1998 UH8K
data with 2000 CFH12K data.  The polynomial coefficients are estimated
under the assumption that the average position of moderately bright
stellar objects has zero motion.  A 3-sigma clipping is applied to
eliminate proper motion stars.  This gives a transform based on 70 to
100 reference stars ($19 < V < 23$) in each CCD. Remote spheroid and
thick disc population dominate reference stars.  With this selection,
we establish from our galaxy model that the zero of proper motions is
roughly moving with a rotation lag of -0.007 arcsec per year along x
and -0.008 arcsec along y with respect to the local standard of rest.

With three epochs it becomes possible to check for the possible existence 
of spurious proper motions which turn out to be 
frequent at faint magnitudes. It is not clear what mechanism
generates such spurious proper motions. The most plausible 
explanation is a displacement of the photocenter of remote
galaxies (not detected as galaxies by the $CLS$ criterion)
 presumably due to supernova events. Their number 
grows rapidly if we release the $CLS$ selection criterion.
 If motions are real, discrepancies between proper motions 
obtained between epoch 2 and epoch 1 on the one hand and between epochs 3
and 2 on the other hand  must be consistent
with each other within $3\sigma$ of the point source position.
 Combining this criterion with a rather 
loose $CLS$ selection threshold ( $CLS  > 0.4 $ at all epochs)
provides bona fide high proper motion stars with minimum loss
of detection efficiency.

Proper motions  are first computed in x and y (the y axis is along  
CCD columns).  The 1998 alignment is chosen as reference.  
In this alignment y is almost exactly the declination axis.   
The $\mu_x$ and $\mu_y$ components are then rotated to the ($\mu_U$,  
$\mu_V$) axes (galactic radius, galactic rotation). 

The magnitude-dependent standard error of star-like objects
is estimated following the same process as in the case of the VIRMOS
field. The trend of the proper motion s.e. with
magnitude is suitably approximated 
by the following formula (in arcsec cen$^{-1}$) :
 
\begin{equation}
  \sigma_{\mu} = 0.74 + \exp(1.58*I_{mag}-37.5)
\end{equation}  

Eventually the proper motion limit for searching halo white dwarfs
was set to  $ 3 \sigma_{\mu} $

%
\subsection{Photometry} 
 
As mentioned in the astrometry subsection, 
The DAOPHOT/ALLSTAR package was used to estimate 
magnitudes together with (x,y) positions on each CCD, in each band, at each epoch. Photometry is performed by fitting a variable empirical 
PSF separately in each one of the CCDs of each mosaic.   
 
\subsubsection{Random errors}  
 
Statistical investigations (not reported here) 
 of the photometric accuracy for stellar objects by comparing several  
 exposures of identical fields at the same 
epoch suggest the following photometric error laws:  

\noindent
  VIRMOS data :
\begin{eqnarray}  
 \sigma_{I} =  0.02+ \exp(- 27 +  I)  \\
 \sigma_{V} =  0.02+ \exp(- 27 +  V)    
\end{eqnarray}  
  SA57 data :
\begin{eqnarray}     
\sigma_{I} =   0.03+ \exp(- 26.1 + 0.958\ I)  \\  
\sigma_{V} =   0.03+ \exp(- 26.6 + 0.934\ V)  
\end{eqnarray}
  
\subsubsection{Systematics}

\begin{figure} 
\epsfig{file=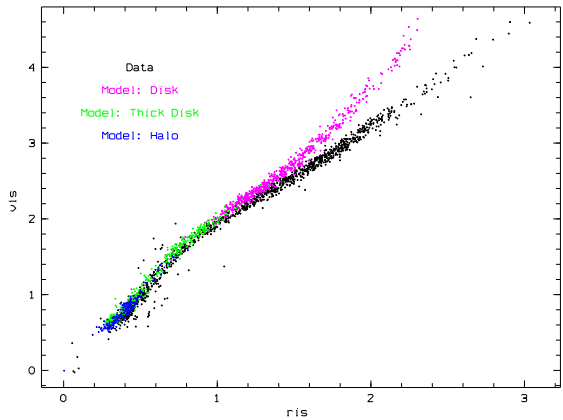,width=6cm,angle=270} 
\caption {(R-I) vs (V-I) for data and model in the VIRMOS field. The plot is limited to bright stellar objects : $CLS > 0.95$, $18<I<20$. Colours are
  black : data , blue : model(spheroid),
green : (thick disc), magenta : (disc) 
 } 
\label{rivi}
\end{figure}

Although the observed photometric data are in principle  
tied to the standard Johnson-Cousins system, there is an obvious 
deviation of observed to theoretical colour-colour diagrams 
(the pattern is identical in both surveys) 
as can be seen in Fig.~\ref{rivi}
 
 
For the purpose of model-data comparisons, model colours were shifted
in a way that forces the model sequence  to fit the real data sequence. 
Assuming that : 
 
\begin{itemize} 
 
\item the shoulder point $(R-I)_{0} , (V-I)_{0} $ at the junction between  
thick disc and disc is a fixed point. 
 
\item   ends of sequences in data and model correspond to similar objects. 
 
\end{itemize} 
  
Colour conversions are then fitted so that the whole model sequence 
is brought into coincidence with the model. 
 
In this way we expect to produce model predicted colours which may 
serve in interpreting the physical properties of real stars at least in a 
statistical way. This rather harsh procedure has only limited consequences  
on the present investigation since the change in $V-I$ is quite small. 
Anyway, there is almost no effect on the blue side, so our results concerning the photometry of ancient white dwarfs is essentially unaffected.


%
%

 
\subsection{ Detection efficiency }

Numerical experiments are used for testing our capacity of identifying 
real stars out of these data.  We seed the original frames with fake 
stellar images built on the observed PSF. The tool for this simulation 
is the ADDSTAR program of the DAOPHOT package. 
 Then images are reprocessed 
and analysed through the complete chain. And detection efficiency is 
established  as the fraction of fake stars seeded into the images 
eventually retrieved and selected at a given CLS level. 
 
\begin{figure} 
 
\epsfig{file=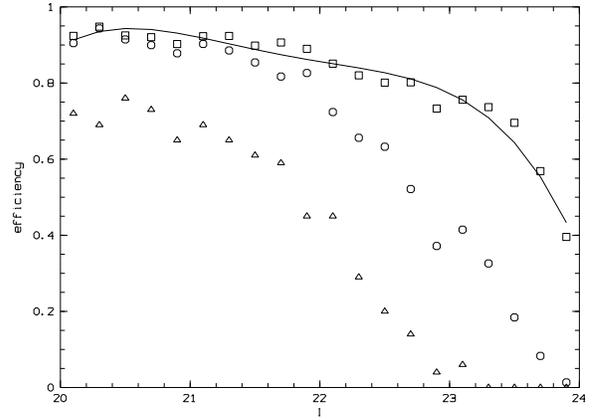,width=6cm,angle=270} 
\caption { Efficiency versus I magnitude under various CLS thresholds :
0.95 (triangles), 0.80 (circles), 0.40 (squares and fitted line)) } 
\label{effi} 
\end{figure}

Up to 2700 stellar images with I magnitude ranging between 20 and 24 have  
been seeded into 9 different CCD frames distributed over the whole VIRMOS  
field at two epochs. Fig.~\ref{effi} shows the fraction of positive 
detections. 
 
 
Efficiencies at CLS threshold 0.4 as approximated by the regression  
curve appearing on the figure are used in section 3 to establish  
the probability for any model predicted halo white dwarf to be detected  
in this survey.

%

\section{ Halo white dwarfs } 
 
%
        %

%
 
A compromise had to be found between adopting a very high $CLS$ 
selection threshold which would warrant almost certainly stellar candidates at the expense of very low efficiency at faint magnitudes  
and low threshold which, at faint magnitudes, would let the sample be 
dominated by non-stellar objects with the prospect of some of them 
producing artifactual proper motions resulting from extragalactic source variability. 
As a result of the detection efficiency study (section 2.6), and of the 
CLS/magnitude statistics (Fig.~\ref{cls-mag}), it turns 
out that a $CLS > 0.4$ at two epochs achieves a minimum loss of 
efficiency while letting in only a small fraction of obvious extended sources 
  
Above magnitude $ I=22.5$ the only way to eliminate galaxies 
without eliminating most stellar objects is offered by proper motions, 
yet at faint magnitudes some artifactual proper motions appear, most related to 
uncorrected edge defects others to pixel defects, 
 some possibly due to supernovae which would displace the photocenter at one epoch. 
Since we have only two epochs in the VIRMOS field there is no way to check 
for the later, while the former can occasionally be disregarded 
 on the basis of a careful examination of images.

\subsection {Implementing a high proper motion limit 
                        in data}

%

\begin{figure} 
\epsfig{file=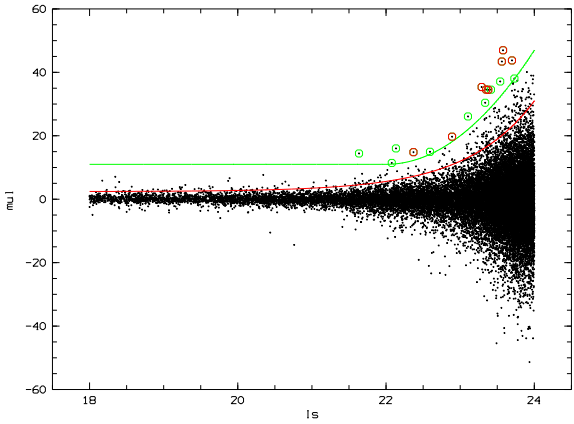,width=6cm,angle=270} 
\caption {Proper motion in longitude versus $I$ magnitude ,
        observed data (VIRMOS field). The green limit is the upper
        envelope of possibly significant proper motions. Below this limit
        proper motions due to unrecognised non-stellar sources may
        contaminate. Circled dots are visually checked objects . All are
        identified as plate defects or misclassified galaxies.The red
      line is the $3\sigma$ limit of the proper motion errors as
      estimated from Table 1} 
\label{wdsearch}
\end{figure} 

Fig.~\ref{wdsearch}  is a plot of $\mu_l$ proper motion  against I  for  
selected stellar candidates in the VIRMOS field. The zero of proper motions corresponds 
to the local standard of rest. Due to the rotation lag, the majority of
nearby halo stars are expected to appear in the upper half of 
this diagram.  
                    
Then, we establish an empirical limit 
( in green on the $ \mu_l $ versus I plot) which is the upper envelope 
of  proper motions in galactic longitude. 
There are seventeen objects found slightly above this line. 
Careful examination on CCD images of these objects above the limit 
shows that all of them correspond to image defects on one 
epoch and/or elongated image. 

The four faintest borderline objects appearing in the VIRMOS
field are too faint for clear discrimination on available CFHT images.
They were submitted for spectroscopic observations 
at the VLT. Actually, preliminary VLT identification frames, 
used as a third observation epoch, were sufficient to
 establish that the high proper motions were all spurious.
 
So there are no proper motion objects detectable above the green 
limit. As can be seen this limit is far above the 3 sigma
 limit of proper motion random errors as derived from
point-like sources (in red ). 
 
\begin{figure} 
\epsfig{file=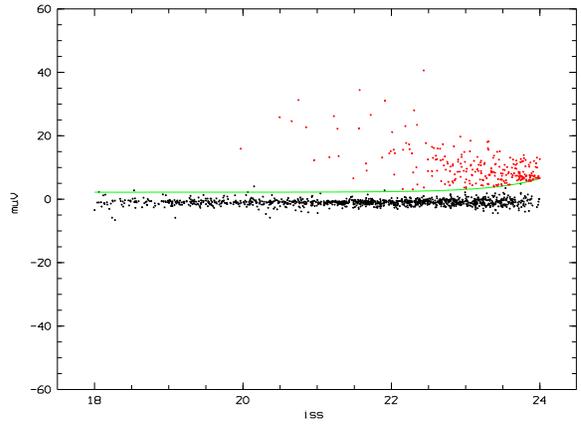,width=6cm,angle=270} 
\caption {Proper motion component along galaxy rotation versus $I$ magnitude.
        Observed data in Sa57 field (black dots). The green limit is the 
        $ 3\sigma $ of pointlike source proper motions. Red dots 
        are halo white dwarfs from model 12 Gyr, IMF2 predicted above
        this limit.}
\label{wdsearch57}
\end{figure}

Data in the SA57 field (Fig.~\ref{wdsearch57}) 
are much easier to
interpret : With three observation epochs, spurious proper motions 
of undiscriminated non-stellar sources are efficiently detected.
The green limit is simply the $ 3\sigma $ limit of stellar 
source proper motions. There are a few real large proper motions
 at magnitudes brighter than $ I=22 $. The magnitudes, colours and
 proper motions of these stars are well compatible with being members 
of standard disc, thick disc and spheroid populations. Model-predicted
halo white dwarfs of similar magnitudes would have much
larger proper motions.

%

%

%

%
                         
\subsection {Models for dark halo white dwarfs} 
 
Four different models of hydrogen halo white dwarf  
populations have been generated assuming halo ages of 12 
or 14 Gyr. White dwarfs are created assuming two different 
initial mass functions and moved to present day colours  
following cooling lines by \cite{Chabrier2000a}. These models do
include H2 collisionally-induced absorption in the atmosphere,
based on the atmosphere models of \cite{Saumon2000}

The parameters of the dark halo adopted for simulation purposes are  
according to \cite{robin2003}. The density law is a power law with an exponent of 2.44 and a flattening of
0.75. The number of simulated white dwarfs should not depend too much
on these parameters since they are detected only at small distances where the
density gradient is rather small. The resulting local mass density is
 $0.0099\ \msun$ pc$^{-3}$. 

The kinematics of the dark halo are assumed identical with those of the
stellar halo: no rotation, ($\sigma_U/\sigma_V/\sigma_W$) = (131/106/85)
km s$^{-1}$.

Models have been generated using
 two initial mass functions according to 
\cite{Chabrier99}: IMF1 has white dwarf mass  $0.8\ \msun$ and IMF2
has $0.7\ \msun$.
For both IMFs we got the resulting white dwarf luminosity function 
after $12$ or $14$ Gyr from \cite{Chabrier2002}.
Actually, the luminosity distribution of the halo white dwarfs
observable in our survey peaks around $M_v = 16.3 $ for IMF2 after
$12$ Gyr and goes as faint as $M_v = 17.3 $ for IMF1 after
$14$ Gyr. Most colours range within $ 1.0\ <\ V-I\ <\ 1.5$.
At such colours, the magnitude limit in the V filter is not an issue.

\begin{figure} 
\epsfig{file=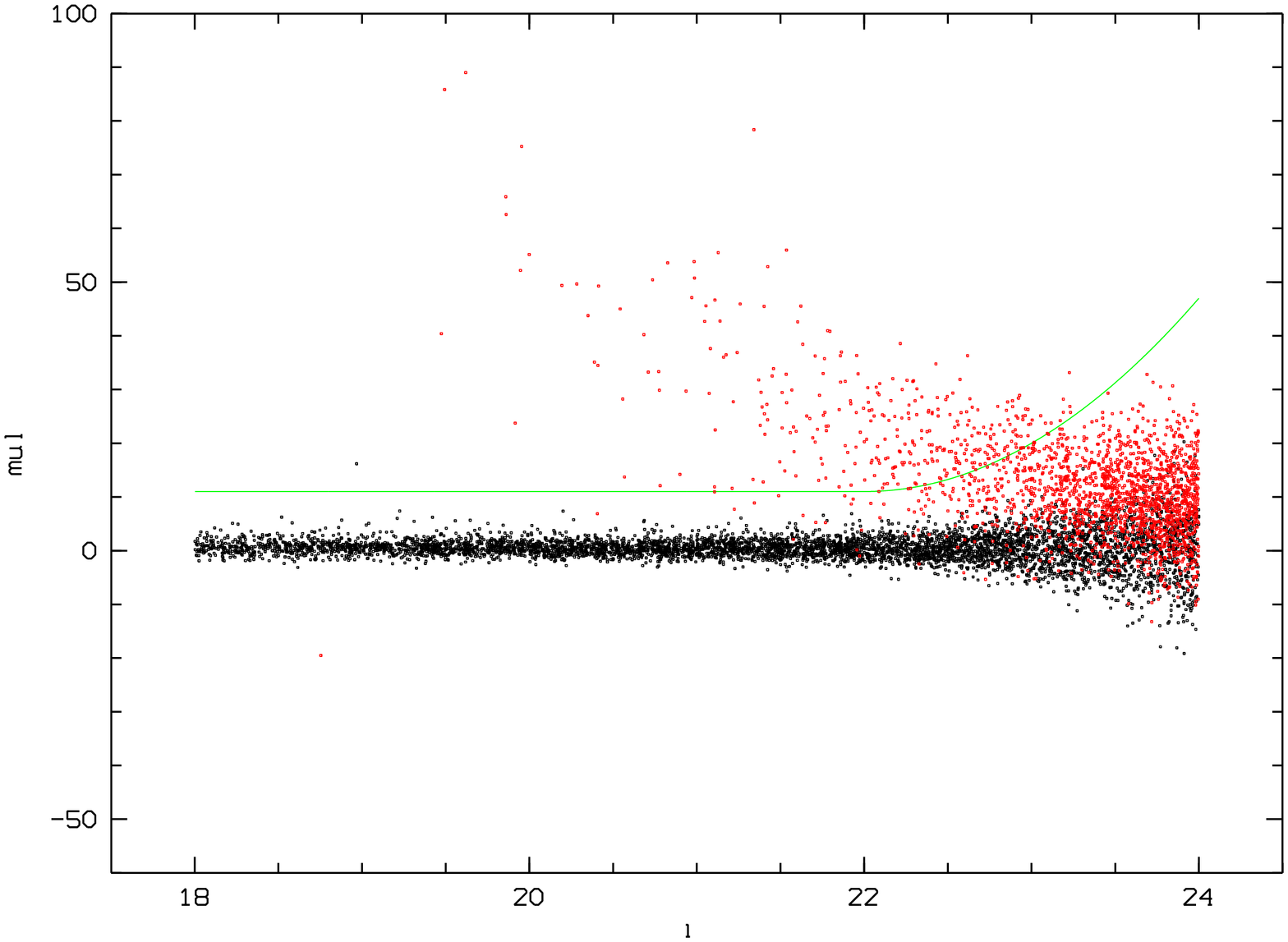,width=6cm,angle=270} 
\epsfig{file=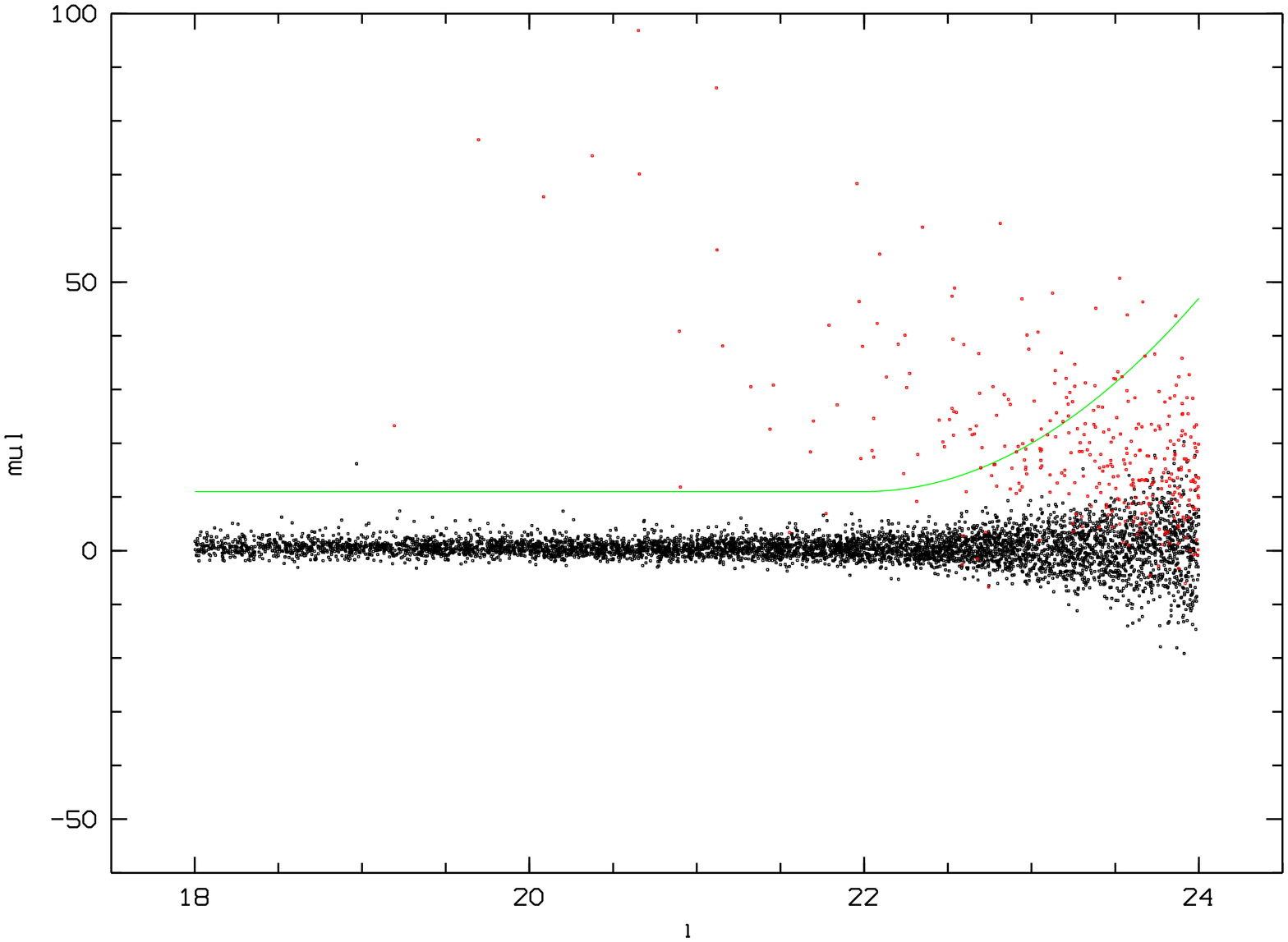,width=6cm,angle=270} 
\caption {I vs. $\mu_l$ plot for simulated data in the VIRMOS field.
 Halo white dwarfs in red. Top : halo age $12$ Gyr , white dwarf
mass $0.7\msun$. Bottom $14$ Gyr, $0.8\msun $ } 
\label{mulI}
\end{figure}

Monte Carlo simulations of dark halos fully made of white dwarfs
according to the above mass and luminosity scenarios have been produced for
magnitude, colour and proper motion observations in our two fields.
The resulting proper motion versus I plots for two extreme 
models are given in Fig.~\ref{mulI}. 
They are over plotted on the
distribution of the three standard populations as predicted by the
Besan\c{c}on model. The upper envelope of observed proper motions as
established in section 3.1 and Fig.~\ref{wdsearch} is also over plotted
in green. The constraint
brought by not detecting proper motion objects above this envelope
is now estimated by simply counting the red points above.

%
 
\subsection {Estimating dark halo fraction} 
 
The number of model predicted halo white dwarfs expected in each field
 in the magnitude,
proper motion volume probed by our survey, above the limits discussed
in section 3.1,  is given in Table~\ref{detect} for each halo
hypothesis. Simulations were performed 
in enlarged fields to secure sufficient statistics. The number of
expected detections used in Table~\ref{hfstat} is obtained by scaling
the numbers to the actual observation field, then applying the average
efficiency factor for this simulation. The average efficiency factors 
are obtained by assigning each star an efficiency loss factor
corresponding to its $I$ magnitude as from the efficiency calibration
given in Fig.~\ref{effi}, then averaging this factor over the stars actually
predicted by the adopted model inside the observation limits.

\begin{table}  
      \caption[]{Number of simulated halo white dwarfs above the green
limits in Fig's \ref{wdsearch57} and \ref{mulI}
under different modelling hypotheses in both observing
fields. Average detection efficiencies are also given. Model labels indicate 
the adopted halo age (first number), and IMF (1) or (2) (see text).}
         \label{detect} 

\begin{center} 
\begin{tabular}{c r r r r r} 
\hline\hline 
\noalign{\smallskip}         
model &$N_\mathrm{vir}$ & $<$ef$_\mathrm{vir}>$ &$ N_\mathrm{SA57}$ &
    $<$ef$_\mathrm{SA57}>$\\ 
\noalign{\smallskip}
\hline 
\noalign{\smallskip}       
      12(1)  &   33.6  &   0.84  &    18.5   &  0.67   \\ 
      12(2)  &   44.8  &   0.85  &    25.8   &  0.68   \\ 
      14(1)  &   13.1  &   0.80  &     6.3   &  0.66  \\ 
      14(2)  &   22.4  &   0.82  &    11.0   &  0.65   \\  
\noalign{\smallskip}
\hline 
\end{tabular} 
\end{center} 
\end{table}

 The probability $p$ of detecting no candidates in given observing
 limits when $m$ detections are expected is $ \exp(-m)$ (Poisson
 distribution).

So if the number of expected detections is $N_d$,
the maximum percent of the halo fraction (hfr) acceptable under odd
probability  $p$ is~:

\begin{equation}  
 \mathrm{hfr}(p,N_d) = (100 * \ln (1/p))/N_d)
\end{equation}

Table~\ref{hfstat} gives the resulting halo fraction under odd probabilities  
 $p=0.05$ and $p=0.36 $ i.e. confidence levels $95\%$ and
 $64\%$ respectively.

Expectedly the estimation of halo fraction is model-dependent.  
The adopted halo age plays the major role, but it also depends on 
the adopted initial mass function. The limit of the halo age is now 
well fixed by CMB fluctuation data at about $13.7$ gigayears. Thus
using figures  
based on 14 gigayears sets the safest limit. The result turns out to be
robust to possible overestimation of efficiencies at faint magnitudes. 

\begin{table}  
\caption{Maximum halo fraction (percent) in the form of ancient white dwarfs
  compatible with zero effective detection,
  given the number of expected detections ($Nd_{...}$). Numbers are
  now scaled to actual field size. Halo fractions are given at 
$95\%$ and $64\%$ confidence levels.} 
\label{hfstat}  
\begin{center} 
\begin{tabular}{c r r r r r}
\hline\hline 
\noalign{\smallskip}
  model & $Nd_\mathrm{vir}$ & $ Nd_\mathrm{Sa57}$ & $ Nd_\mathrm{tot}$
  &  hfr$_{95}$ & hfr$_{64}$\\ 
\noalign{\smallskip} 
  \hline
\noalign{\smallskip} 
  12(1) &  28.29 &  12.40 & 40.69 &   7.4 &   2.5 \\
  12(2) &  37.95 &  17.54 & 55.49 &   5.4 &   1.8 \\
  14(1) &  10.51 &   4.16 & 14.66 &  20.4 &   7.0 \\
  14(2) &  18.41 &   7.15 & 25.56 &  11.7 &   4.0 \\
\noalign{\smallskip}
  \hline 
\end{tabular} 
\end{center} 

\end{table}  
 
%
 
\section{Comparison with other investigations} 
 
%
%
 
Similar investigations have been published by  
\cite{goldman} based on a specific proper motion 
 survey by the EROS group within $250\sqd $ down to $ V=21.5$, and
by \cite{nelson} based on a search for proper motion stars in HST
 frames in a much smaller 
field $( 0.02078\sqd)$ yet much deeper $( V \le 26.5, V-I \le 1.62)$. 
In the following we attempt to reinvestigate the two data sets 
using the approach we had with our own data. That is :
identify a volume where halo white dwarfs if they exist would be the
only significant contributors to the statistics. In the conclusions we
will combine the three results.

\subsection{EROS white dwarf search} 
The EROS group conclusion is that they exclude halo mass fractions 
larger than $5\%$ in the form of ancient white dwarfs. The MACHO/HST 
group concludes that some of the white dwarfs they have detected `may 
be members of the halo'. In order to combine these results with our 
own, we re-evaluate the constraint brought by each data set in an 
homogeneous way. That is, based on our modelling of known stellar 
populations, we delimitate a region of the magnitude proper motion 
 space where no stars are expected except for halo white dwarfs. And no 
 candidate is found. 
 
Then we  compute our own model prediction for a hundred percent white 
 dwarf halo within this limit, since there is no detection in any of 
 the three samples. 
 
In \cite{goldman} the limits of the zero candidate volume are
 defined as : 

\begin{eqnarray} 
  70 \mathrm{arcsec cen}^{-1} < \mu < 600 \mathrm{arcsec cen}^{-1}  \\ 
  V < 22 \\ 
  H_V > 22.5\ ;\ H_V = V + 5 \log (\mu/100) + 5 
\end{eqnarray}
 where 
\begin{equation} 
  H_V = V + 5 \log (\mu/100) + 5 
\end{equation}

The detection efficiencies have been estimated by Goldman (private 
communication). It degrades from $0.92$ at magnitude $16$ down to $0.29$  
at magnitude $21$. Following the same procedure as in the case of our
own data each predicted star is assigned a weight equal to the 
efficiency corresponding to its magnitude. Then the number of 
expected detections is simply the sum of the weights of predicted  
white dwarfs.  
 
\subsection {MACHO/HST white dwarf data} 

 \cite{nelson} used two epoch exposures of the Wide Field Planetary
 Camera in the so-called Groth-Wesphal strip. In this 74.8 arcmin$^2$
 survey complete to $V\ =\ 26.5$ they identify a series of high proper
 motion objects out of which $5$ are considered as strong white dwarf
 candidates. That is, according to the authors, about three more than
 expected from standard disc, thick disc and spheroid
 populations. They tentatively interpret their findings
 as follows : \textit{ `Possibly, the excess cannot be
 explained without invoking a fourth galactic component'} actually a
 white dwarf dark halo. Yet they admit that this conclusion is rather
 model-dependent. 
\begin{figure} 
\epsfig{file=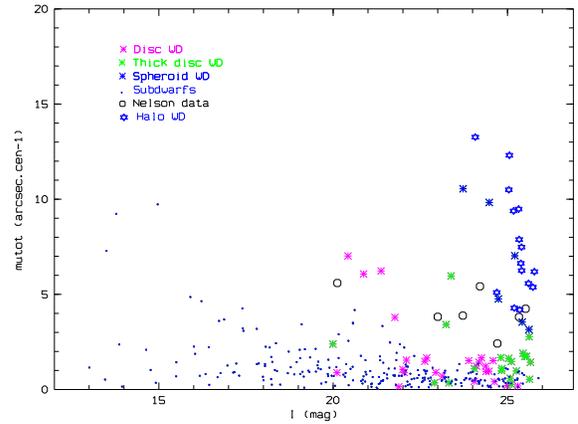,width=6cm,angle=270} 
\caption { $|\mu|$ versus $I$ distribution of the HST
observations \cite{nelson} compared to model predictions.
 Model predictions for standard populations are amplified
by $10$ 
 } 
\label{nelsonfig}
\end{figure} 

Indeed our simulations for the HST field (Fig.~\ref{nelsonfig}) confirm
that some of their 
observed white dwarfs have  magnitudes and proper motions
compatible with being halo members (They are also compatible
with being thick disc or 
spheroid members). We also confirm that the detected objects outnumber
 what can be accounted for from standard populations : It is necessary to
 amplify white dwarf model predictions by at least a factor of 5 to
produce a number of thick disc and/or spheroid white dwarf members.

But our simulation shows also that any white dwarf halo model capable
of placing 
a few stars at this place in the magnitude/proper motion diagram
would at the same time produce many more detectable stars with  higher 
proper motions where there would be no possible confusion with standard
populations. Among the models we tried, those capable of
producing one halo 
white dwarf with $I<25.5$ and total proper motion smaller than $6$
arcsecond per century produce at least 10 proper motions larger
than this limit where Nelson's data has none. So the interpretation
of Nelson's stars as halo white dwarf is quite unlikely.

We infer from the \cite{nelson} observations that in a volume within
which a full white dwarf halo would put 9 or 10 objects, they detect 
none. These figures will be combined with others in the next section.

One should also remember that another search based on very deep
HST observations produced negative results \citep{ibata, richer2002}.

In Table~\ref{hftot} we replicate the estimation presented in
section 3.3 incorporating the additional constraints from \cite{nelson}
and  \cite{goldman}.

\begin{table}
\caption{Same as Table~\ref{hfstat} combining data analysed 
in this paper( Nd$_\mathrm{C}$), Nelson's (Nd$_\mathrm{N}$) and Goldman's (Nd$_\mathrm{G}$).
 Models are 14.1 (top)and 14.2 (bottom)}  
\label{hftot}
\begin{center} 
\begin{tabular}{r r r r r r}
\hline\hline 
\noalign{\smallskip}
Nd$_\mathrm{C}$ & Nd$_\mathrm{N}$ & Nd$_\mathrm{G}$ &
Nd$_\mathrm{tot}$ & hf$_\mathrm{95}$ & hf$_\mathrm{64}$ \\  
\noalign{\smallskip}
\hline
\noalign{\smallskip}
 14.7 &   9 &  26.8 &  49.3 &  6.1 &  2.1 \\ 
 25.6 &  10 &  46.8 &  80.0 &  3.7 &  1.3 \\ 
\noalign{\smallskip}
\hline 
\end{tabular} 
\end{center}
\end{table} 
 
With EROS data alone \citep{goldman}  this estimation would have produced a  
maximum halo fraction of $11.2\%$ ($0.95$ confidence level) under IMF1
 and  $6.4\%$ under IMF2. With the same data and an IMF close to IMF2
the limit obtained by the authors is $5\%$ producing a sharper result.
The discrepancy comes from slightly different IMF and halo
parameters. For the sake 
of robustness, we used always the most conservative hypotheses. Had we
used exactly the same modelisation, it would make the halo fraction
even smaller.
It would be tempting to combine these results also with the constraint
brought by the volumes explored by \cite{flynn2001} using Luyten's LHS
and NLTT proper motion surveys. However the compleness of this survey
at faint magnitudes is not well established beyond Luyten's R
magnitude 18. There is no possibility to directly calibrate detection
efficiency on Luyten's plates and the indirect calibration used
by \cite{flynn2001} is subject to biases. So we can only notice that
constraints brought by Luyten's data strengthen our conclusions
without being able to make it a quantitative statement.

\section{Conclusions}

Three deep proper motion surveys produce zero halo white dwarfs within
observational limits where a full white dwarf halo would exhibit 
between 49 and 80 such objects. 

This is in obvious conflict with the conclusions of two investigations
based on less deep observations
\citep{Oppenheimer, Mendez2002}. Although
exploring comparable or larger galaxy volumes, surveys at brightest
magnitudes cannot safely isolate halo stars. In
 the two quoted references, the claimed halo white dwarfs
have proper motions quite compatible with being thick disc stars.
 The diagnosis of halo has been rejected by many
different sources under different approaches: \cite{Reyle2001b}, \cite{Reid},
\cite{silvestri2002}, \cite{Torres2002}, \cite{flynn2003}. All support the idea
that the thick disc is the most likely explanation.

So the convergence of existing data is quite
clear : whenever ancient halo white dwarfs have been searched for under
 observing conditions where no room was left for confusion with
 existing standard populations, there were none found. 
The combined constraint from existing investigations
pushes the maximum halo fraction below $4 \% $ at the $95\%$ confidence level
and below $1.3 \% $ ($64\%$ confidence). 

 This direct
 observational evidence is consistent with arguments based on
 the overproduction of carbon relative to oxygen that
 would accompany the evolution of halo WD precursors \citep{Brook2003}.
 The only way to keep some limited room open for
this hypothetic contributor to the Dark halo baryons would be to
adopt Chabrier's IMF1, that is to assume that most dark halo white
dwarfs have masses about $0.8 \msun$. Yet according to
\cite{Chabrier99}, the progenitors of such high mass 
white dwarfs would return more processed gas into the interstellar
medium per unit mass of white dwarfs produced. So they would
contribute more efficiently to the early carbon enrichment. 

 This implies that if baryonic dark matter is present in galaxy halos,
 it is not, or is only marginally in the form of faint hydrogen white dwarfs.


%
 
\begin{acknowledgements} 
 
This research work was partially supported by the Indo-French Centre for the 
 Promotion of Advanced Research (IFCPAR) / Centre Franco-Indien pour la 
 Promotion de la Recherche Avanc\'ee (CEFIPRA), New Delhi (India).We
 thank Gilles Chabrier for providing his models in advance of
 publication and for useful advice all along this investigation. We
 also thank David Graff, Bertrand Goldman and an anonymous referee
 for several constructive remarks.

\end{acknowledgements} 
 
%
  
\appendix\

%
 
%

%

\end{document}